\documentclass[default,iicol,sn-aps,twocolumn]{sn-jnl}

\usepackage{graphicx}%
\usepackage{multirow}%
\usepackage{amsmath,amssymb,amsfonts}%
\usepackage{amsthm}%
\usepackage{mathrsfs}%
\usepackage[title]{appendix}%
\usepackage{xcolor}%
\usepackage{textcomp}%
\usepackage{manyfoot}%
\usepackage{booktabs}%
\usepackage{bbding}
\usepackage{algorithm}%
\usepackage{algorithmicx}%
\usepackage{algpseudocode}%
\usepackage{listings}%
\usepackage{physics}
\usepackage{adjustbox}
\usepackage{pifont}

\newtheorem{theorem}{Theorem}
\newtheorem{lemma}{Lemma}

\begin{document}

\title[Article Title]{Hybrid Quantum-Classical Clustering for Preparing a Prior Distribution of Eigenspectrum}

\author[1]{\fnm{Mengzhen} \sur{Ren}}
\equalcont{These authors contributed equally to this work.}
\author*[2]{\fnm{Yu-Cheng} \sur{Chen}}\email{kesson.yc.chen@foxconn.com}
\equalcont{These authors contributed equally to this work.}
\author[3]{\fnm{Ching-Jui} \sur{Lai}}
\author[2]{\fnm{Min-Hsiu} \sur{Hsieh}}
\author*[1,4]{\fnm{Alice} \sur{Hu}}\email{alicehu@cityu.edu.hk}

\affil[1]{\orgdiv{Department of Mechanical Engineering}, \orgname{City University of Hong Kong}, \orgaddress{\city{Hong Kong SAR}, \postcode{999077}, \country{China}}}

\affil[2]{\orgname{Hon Hai (Foxconn) Research Institute, Taipei, Taiwan}, \orgaddress{\city{Taipei}, \postcode{114}, \country{Taiwan}}}

\affil[3]{\orgdiv{Department of Mathematics}, \orgname{National Cheng Kung University}, \orgaddress{\city{Tainan}, \postcode{70101}, \country{Taiwan}}}

\affil[4]{\orgdiv{Department of Materials Science and Engineering}, \orgname{City University of Hong Kong}, \orgaddress{\city{Hong Kong SAR}, \postcode{999077}, \country{China}}}

\abstract{Determining the energy gap in a quantum many-body system is critical to understanding its behavior and is important in quantum chemistry and condensed matter physics. The challenge of determining the energy gap requires identifying both the excited and ground states of a system. 
In this work, we consider preparing the prior distribution and circuits for the eigenspectrum of time-independent Hamiltonians, which can benefit both classical and quantum algorithms for solving eigenvalue problems. The proposed algorithm unfolds in three strategic steps: Hamiltonian transformation, parameter representation, and classical clustering. These steps are underpinned by two key insights: the use of quantum circuits to approximate the ground state of transformed Hamiltonians and the analysis of parameter representation to distinguish between eigenvectors. The algorithm is showcased through applications to the 1D Heisenberg system and the LiH molecular system, highlighting its potential for both near-term quantum devices and fault-tolerant quantum devices. The paper also explores the scalability of the method and its performance across various settings, setting the stage for more resource-efficient quantum computations that are both accurate and fast. The findings presented here mark a new insight into hybrid algorithms, offering a pathway to overcoming current computational challenges.}


\maketitle
\section*{Introduction}\label{sec1}
The energy gap lies at the heart of understanding the quantum many-body system, providing a view into many important physical and chemical behaviors \cite{sugisaki2023projective, mcardle2020quantum}. In quantum chemistry, the energy gap is deemed as the stability indicator of molecules and is also related to the activation energy required for many chemical reactions. Another notable application is the energy gap between the Highest Occupied Molecular Orbital (HOMO) and the Lowest Unoccupied Molecular Orbital (LUMO) called the HOMO-LUMO gap \cite{shee2023quantum}. This gap helps predict the robustness and stability of the transition metal complex \cite{aihara1999reduced}. For the quantum many-body systems, the spectral gap, defined as the difference between the two smallest eigenvalues, is a crucial parameter in comprehending the ground space of any Hamiltonian. It further unveils information about phase transitions and the characteristics of quantum phases \cite{cubitt2015undecidability}. In materials science, energy bands describe the allowed energies of electrons in a crystal lattice, the band gaps between these bands determine the material's electrical and optical properties and serve as a key characteristic for understanding the behavior of material properties \cite{crowley2016resolution, cerasoli2020quantum}.

Determining the specific energy gap requires the resolution of the system’s eigenspectrum in the vicinity of the gap. However, solving exact eigenvalues in complex systems presents a formidable challenge, seemingly necessitating exponential time for both classical and quantum computation \cite{osborne2012,kempe2006}. Even ascertaining the gapped or gapless of the Hamiltonian is an undecidable problem \cite{cubitt2015undecidability}. Therefore, emphasis is shifted to the development of more efficient methods for approximating the eigenspectrum around the target energy region. The refinement promises to be more practical and achievable in exploring energy gaps.

To approximately solve the eigenspectrum, knowing prior knowledge of the eigenspectrum plays an important role. The classical iterative algorithms such as the power, Krylov subspace, Lanczos, and Arnoldi methods and their variant \cite{young2014iterative,lanczos1950iteration,arnoldi1951principle}, are adept at handling sparse matrices often encountered in Hamiltonian problems. Although they may struggle with large and complex systems in terms of computation and accuracy, understanding prior knowledge of the eigenspectrum can allow for targeted computations that reduce overall complexity. On the other hand, the Fault-Tolerant Quantum methods such as the Phase Estimation algorithm, Harrow-Hassidim-Lloyd (HHL) algorithm, and Quantum Singular Value Transform (QSVT) algorithms leverage quantum computation and algorithms to find approximate solutions when offering exponential speed-up over classical methods for classes of Hamiltonians \cite{kitaev1995,Abrams1997,abrams1999,harrow2009,cao2019,zhang2022quantum}. However, they all require the prior of eigenspectrum like the rough estimation of spectrum range, target eigenvalue, and spectral gap to have the advantages. Moreover, current quantum methods must contend with near-term limitations such as coherence times and error rates, which can weaken their effectiveness. 

Since it is important to obtain the prior eigenspectrum with acceptable accuracy using as few resources as possible, whether there is a quantum advantage in the preparation of prior distribution is an important issue. Adjusting the trade-off between the quality and cost of the solution by varying the number and the precision of the quantum and classical steps \cite{lau2022nisq}, the hybrid algorithms can be deemed as a prior eigensolver for both classical or Fault-Tolerant Quantum methods. Variational methods \cite{aspuru2005,cerezo2021,vqe1,vqe2,vqe3,kao2023exploring,li2023gell}, Quantum Krylov subspace methods (QKS) \cite{krylov1,krylov2,qlanczos,qdavidson,Seki2021}, and Monte Carlo-based methods \cite{yang2024resource} are prime examples that benefit from emerging accuracy of noisy intermediate-scale quantum (NISQ) devices. Although they can use quantum computing to explore the exponentially large Hiber space, solving the prior spectrum is not the original goal of these methods. Therefore, each of them faces different challenges to obtain this prior information efficiently.

The first challenge is the circuit depth problem, methods like Variational Quantum Deflation (VQD) need to compute excited states sequentially, which requires scaling circuit preparation to shift up all low-energy states. On the other hand, QKS and Monte Carlo methods also require non-shallow circuits to prepare real-time evolution states and examine the Rayleigh quotient and state fidelity. The second challenge is the ansatz design problem since Variational methods heavily rely on the optimization of parameterized unitary circuits concerning the objective function. The third challenge is the measurement overhead. Within the framework of Monte Carlo-based methods, a substantial sample size is unavoidable to approximate the underlying distribution accurately. Furthermore, the VQD requires the precise measurement of the overlap between quantum states. The final challenge is the inability to compute higher excited states. Krylov subspace methods suffer from orthogonality loss and can only compute the first few excited states based on the residual criterion. 

The reason for the challenges is that the design of these algorithms is aimed at accuracy. However, to prepare the prior spectrum, the aim of a solver is only to provide an acceptable resolution to the spectrum, and it can give feedback to the existing algorithms to get more accurate results deterministically. In Table \ref{tab:summary} we demonstrate the challenges of preparing prior eigenspectrum using different hybrid algorithms, the \ding{51} implies the method can be free from the challenges.

\vspace{-0em}
\begin{table}[!htp]
\begin{adjustbox}{width=\columnwidth,center}
\begin{tabular}{@{}ccccc@{}}
\toprule
Challenges\textbackslash{}Methods & \begin{tabular}[c]{@{}c@{}}Variational\\  methods\end{tabular} & \begin{tabular}[c]{@{}c@{}}QKS\\methods\end{tabular} & \begin{tabular}[c]{@{}c@{}}Monte Carlo\\ based methods\end{tabular} & \begin{tabular}[c]{@{}c@{}}Our\\ method\end{tabular} \\ \midrule
Scaling Circuit Depth &  &  &  & \ding{51} \\ 
Strict Ansatz Design&  & \ding{51} & \ding{51} &  \ding{51} \\  
Measurement Overhead & \ding{51} & & & \ding{51}  \\ 
\begin{tabular}[c]{@{}c@{}} Limited Energy Spectrum\end{tabular} &  &  & \ding{51} & \ding{51} \\ \bottomrule
\end{tabular}
\end{adjustbox}
\caption{Main challenges for preparing prior eigenspectrum using different hybrid algorithms, \ding{51} implies the method can be free from the challenges.}
\vspace{-0.5em}\label{tab:summary}
\end{table}

In this work, we propose a hybrid method that aims to efficiently prepare the prior distribution of a time-independent Hamiltonian $H$ with non-degenerate eigenvalues $\lambda$ and eigenstates. The algorithm unfolds in three pivotal steps:
\begin{itemize}
    \item Hamiltonian Transformation. Introducing a drift parameter $s$, we transform $H$ into a new Hamiltonian $H(s)$. This transformation establishes a one-to-one correspondence between the eigenvalues and eigenstates of $H$ and $H(s)$, where the ordering of eigenvalues may dynamically change with the variation of $s$.
    \item Parameter Representation. Employing a Quantum Circuit with tunable parameters, we generate trial states approximately representing the ground state of $H(s)$. Systematically varying $s$, we find the parameters to feature the ground state of $H(s)$ for each $s$.
    \item Classical Clustering. A classical clustering algorithm categorizes the collected parameters based on their resemblance. Each cluster is associated with an eigenvector of $H$, and the median of $s$ within a cluster approximates the corresponding eigenvalue.
\end{itemize}

The algorithm presented integrates quantum and classical components to exploit the advantages of both methods. The main advantage is based on two key insights: Theorem \ref{the:fid_req_clu}, utilizes quantum circuits to approximate the ground state of the Hamiltonian $H(s)$, which mitigates the classical hardness of wavefunction representation. The parameters derived from the circuit allow us to distinguish different eigenvalues and eigenvectors by analyzing the circuit parameters instead of the ground state itself. Theorem \ref{the:tau_save}, suggests that classically clustering quantum states represented by their circuit parameters can relax the convergence requirements of the quantum part. This relaxation reduces the total time steps needed to obtain a prior knowledge of eigenspectrum and eigenvectors, thus appropriately reducing the quantum resource requirement.

For the first key insight, depending on the availability of fault-tolerant devices, we can divide the methods of achieving approximating the ground state of the Hamiltonian $H(s)$ using quantum circuits into two categories. For the first category, which is feasible in the NISQ era, we can consider using the squared Hamiltonian $H(s)=(H-sI)^2$, where $I$ is the identity operator. In this case, we can use the imaginary time evolution (ITE) algorithms \cite{mcardle2019variational, tsuchimochi2023improved} to calculate the expected value $f(t)=(B(t)v)^\dagger H (B(t)v)$ of $H(s)$, where $B(t)v=\frac{A(t)v}{||A(t)v||_2}$ and $A(t)=\exp\{-H(s)t\}$. The parameters of parameterized quantum circuits (PQCs) or parameterized unitary we obtain by ITE corresponds to the quantum state that is close to the target eigenstate, i.e., the ground state of $H(s)$. 
In contrast, for the second category, which requires fault-tolerant and error-correcting equipment, we can consider using the inverse Hamiltonian $H(s)=(H-sI)^{-1}$. In this case, we can use the HHL algorithm \cite{harrow2009} or Quantum singular value transformation (QSVT) \cite{gilyen2019quantum} to obtain $H(s)$. The quantum state obtained by variational optimization \cite{chiew2023scalable} or inverse power \cite{kyriienko2020quantum} of the input state is also close to the target eigenstate and can be represented in a polynomial number of parameter representation. In both NISQ and FTQC cases, parameter sets that fulfill the approximation for distinct eigenvectors of $H$ can be separated in neighborhoods, the details are also discussed in the Result section. In addition, in the numerical experiments, we also tested the clusterability of the obtained circuit parameter data according to the Hopkins statistic, and the results were consistent with our point of view.
For the second key insight, clusterability can relax the fidelity requirement to distinguish different eigenstates, which leads to total convergence step reduction, fewer measurements, and resilience to noise. First, to quantify the total convergence step reduction, we integrate the quantum speed limit delineated in Lemma \ref{lemma} with the moderated convergence threshold $\delta$ as stipulated in Theorem \ref{the:fid_req_clu}. Consequently, the total time savings $\tau_{\text{save}}$, achieved in the acquisition of a priori eigenspectra and eigenvectors, is bound by the inequality $\tau_{\text{save}} \geq \frac{4 \delta^2}{\beta\pi^2 \langle H(s) - \sigma_{\text{min}} \rangle}$. For a comprehensive exposition of this relationship, refer to Theorem \ref{the:tau_save}. Then, for the fewer measurements, it can be easily realized that the larger tolerance of fidelity error can separate different eigenstates, which will require fewer measurements to prepare a parameter representation of the desired error. Finally, the noise lead fidelity error can also be mitigated by the existence of clusterability, which we have further investigated in the numerical tests with scaling depolarization error.

Our method has several advantages over existing approaches. The algorithm harnesses the computational prowess of quantum computers to approximate states from uniform initial conditions to those corresponding to various eigenvectors of a Hamiltonian $H$, utilizing parameters as stand-ins for the eigenvectors themselves. This approach allows for compact compression in the analysis of eigenstates, which is particularly beneficial for sparse matrices $H$. Furthermore, we integrate classical clustering techniques to moderate the precision needed for the parameters, thereby considerably diminishing the quantum computational resources required. This dual strategy of state approximation through parameterization and resource optimization via clustering positions our algorithm has practical implications for overcoming current computational limitations.

We demonstrate the effectiveness of our method on two examples: the 1D Heisenberg system and the LiH molecular system. We also discuss the potential applications of our method for near-term quantum devices. Moreover, we show how our method can be extended to use the HHL algorithm when fault-tolerant quantum computers are available. Finally, we perform a numerical study on the scalability of our method and analyze its performance under different settings.

The rest of the paper is organized as follows. We delineate the principal components of the algorithm, establishing the clusterability of circuit parameter data as a prerequisite for the quantum circuit ansatz, which corresponds to the quantum state under investigation. We then examine the feasibility, potential benefits, and inherent complexity of the algorithm on NISQ devices and Fault-Tolerant Quantum Computing (FTQC) systems. Subsequently, we demonstrate the application of our algorithm across various models and case studies of differing magnitudes. Finally, we reflect on our findings and propose potential avenues for future research.

\section*{Result}\label{sec2}
We begin by introducing the necessary notation and presenting the algorithmic framework. Our primary objective is to obtain approximate energy spectra and their corresponding quantum states efficiently.

\begin{figure}[!htp]%
\centering
\includegraphics[width=0.5\textwidth]{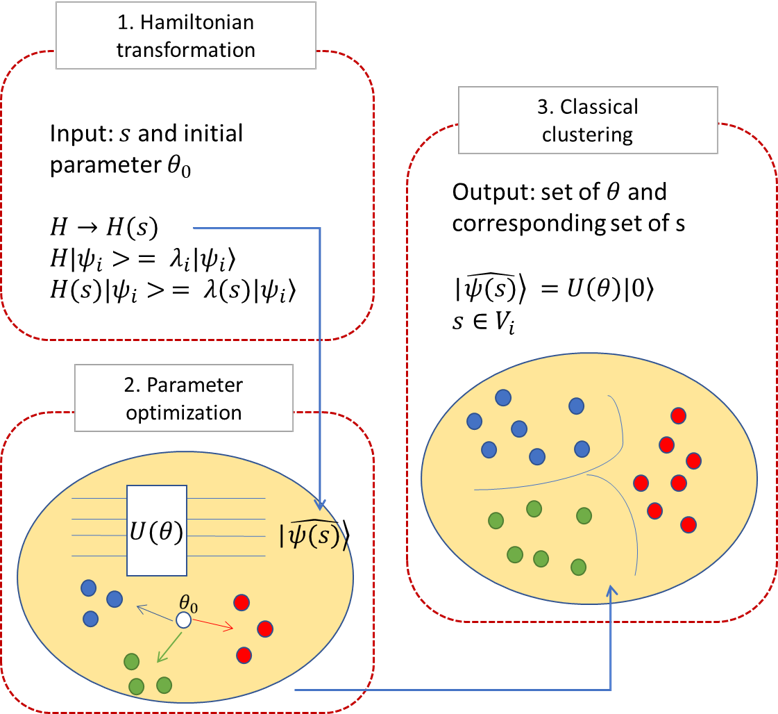}
\caption{Framework of our model.}\label{fig1}
\end{figure}
Firstly, we uniformly select a drift parameter, denoted as $s$, from the interval of interest $V$. For instance, $V$ may represent the value range of the entire eigenspectrum. Transform the original Hamiltonian, denoted as $H$, into a modified Hamiltonian $H(s)$ based on the chosen drift parameter. This transformation allows us to incorporate the effects of the selected drift. Next, the ground state $\psi(s)$ of $H(s)$ is then mapped to the parameters of a quantum circuit, denoted as $U(\theta)$. This mapping enables the core operation of our model. Finally, we estimate relevant quantities based on the obtained parameters of the quantum circuit.

After obtaining the interval $V_i$ of $s$ under the corresponding category through classical clustering of parameters, we obtain the value interval $V_i$ of the corresponding eigenvalue $\lambda_i$. Notably, if we can drive every Hamiltonian operator $H(s)$ to its ground state (e.g., considering $H(s) = (H-sI)^2$), the resulting value range for $V_i$ is denoted as $[\frac{\lambda_{i-1}+\lambda_i}{2},\frac{\lambda_{i+1}+\lambda_i}{2}]$. It seems that if the median or average is used to represent the estimate of the eigenvalue $\lambda_i$, the error will be more obviously affected by the values of nearby eigenvalues and the value range of $V_i$ will be larger. In fact, our proposed method emphasizes the use of shorter quantum circuits and iterative steps. In such cases, the value range of $V_i \subset [\frac{\lambda_{i-1}+\lambda_i}{2},\frac{\lambda_{i+1}+\lambda_i}{2}]$ and $\lambda_i \in V_i$. Consequently, using the median or average to represent the estimated characteristic value $\lambda_i$ yields smaller errors.  For instance, consider the VITE algorithm we used: when $s$ and one of the eigenvalues of $H$ are closely spaced, convergence occurs rapidly. Conversely, if $s$ lies near the average of any two adjacent eigenvalues of $H$, convergence slows down (see the Method for details).

\subsection*{Parameter representation} \label{subsec2.1}
In this section, we delve into the properties of quantum circuit parameters. Specifically, we focus on the parameter set that satisfies the eigenvector fidelity requirements of the Hamiltonian operator. This parameter set plays a crucial role in clustering algorithms. Additionally, we explore the limit of the time required $\tau$ to transform a quantum system while adhering to these fidelity requirements. Our analysis provides a foundational insight supporting the feasibility and advantages of our proposed algorithm.

Although a formal definition of clusterability is lacking, according to the reference \cite{adolfsson2019cluster}, we observe that when dealing with a dataset containing multiple clusters, there should be discernible separation between these clusters. Conversely, homogeneous data should not exhibit such distinct separation properties. Based on this, we explore whether parameter sets satisfying different eigenvector approximation requirements of $H$ can be separated in the neighborhood. If so, this means that the set of quantum circuit parameters obtained according to our requirements should have a certain clusterability. In the Numerical experiments section, we combined the Hopkins statistic to numerically verify the clusterability of the parameter set. We begin our account of the properties of quantum circuit parameters by presenting the following theorem.

Let $H(s)$ be a Hamiltonian operator with $s$ as a drift parameter, and let $f$ denote the evolution of $H(s)$. The following properties are established:\\
\begin{itemize}
    \item The evolution $f$ converges to the eigenvalue of $H$ closest in absolute distance to the drift $s$.
    \item The state $\psi(s)$, resulting from the evolution of $H(s)$ in time $t$, converges to the corresponding eigenstate of $H$ as $s\to0$. 
    \item The corresponding approximating state of $\psi(s)$ in a set of parameters is $\widehat{\psi}_s(\theta)$, where $\theta=(\theta_1,\ldots,\theta_N)$.
\end{itemize}
Define the sets:\\
\begin{itemize}
    \item $S = \bigcup_{i=1}^m S_i$, where $S_i = \{s \mid \lambda_i = \arg\min_{\lambda_j} |\lambda_j - s|\}.$
    \item $\Psi = \bigcup_{i=1}^m \Psi_i$, where $\Psi_i = \{\psi(s) \mid s \in S_i\}$.
    \item The set $\widehat{\Psi}_i$ corresponds to the set $\Psi_i$ of state $\psi(s)$.
\end{itemize}

\begin{theorem}\label{the:fid_req_clu}
(Clusterability of Parameter Representation)
Given $\epsilon_1$, $\epsilon_2$, and $\delta$ such that $\frac{\pi}{2} > \epsilon_2 >2\delta> \delta > \epsilon_1 > 0$. Suppose the following conditions hold: For any $\psi_1 \in \Psi_i$, $\psi_2 \in \Psi_j$, and $s \in S_i$, $(1)$ $\| \psi_1-\psi_2 \|_F < \epsilon_1$ whenever $i=j$, $(2)$ $\| \psi_1-\psi_2 \|_F > \epsilon_2$ whenever $i\neq j$ $(3)$ $\| \widehat{\psi}_s(\theta) - \psi \|_F < \delta$ for all $\psi \in \Psi_i$. 
Then for $D_i( \delta) = \{\theta \mid \|\widehat{\psi}_i(\theta) - \psi_i \|_F < \delta,\ \forall\ \widehat{\psi}_i(\theta) \in \widehat{\Psi}_i,\ \psi \in \Psi_i\}$, the sets $D_i(\delta)$ and $D_j(\delta)$ are disjoint for all $i \neq j$.
\end{theorem}
\noindent\textit{Proof}: For any $\widehat{\psi}_i(\theta) \in \widehat{\Psi}_i$ and $\psi_j \in \Psi_j$, we have:
\begin{itemize}
    \item if $i=j$,\begin{equation}\label{s:i=j}
        \| \widehat{\psi}_i(\theta)-\psi_j\|_F < \delta;
    \end{equation}
    \item if $i\neq j$,\begin{equation}\label{s:i!=j}
        \|\widehat{\psi}_i(\theta)-\psi_j\|_F\geq\left|\|\widehat{\psi}_i(\theta)-\psi_i\|_F-\|\psi_i-\psi_j\|_F\right|>\delta.
    \end{equation}
\end{itemize}
This implies that $D_i(\delta) \cap D_j(\delta) = \varnothing$ for all $i \neq j$.

Since $\|\widehat{\psi}_i(\theta) - \psi_j\|_F$ is a continuous function in $\theta \in [0,T_l]^N$, for any $\boldsymbol{\theta} \in D_i(\delta)$ and $\boldsymbol{\theta}' \in D_j(\delta)$, there exist closed neighborhoods $B_i$ of $\boldsymbol{\theta}$ and $B_j$ of $\boldsymbol{\theta}'$ such that $B_i$ and $B_j$ are disjoint. Therefore, the sets $D_i(\delta)$ and $D_j(\delta)$ are separated by neighborhoods for all $i \neq j$. $\hfill\square$

For instance, if we perform the VITE algorithm from a fixed initial point, which means that we can only obtain the neighborhood of a few or only one locally optimal solution, it is possible to classify the parameter data using some clustering methods.

Next, in assessing the temporal efficiency gained in quantum evolution by moderating the convergence criteria in line with the parameter set's clustering standard, we invoke the Margolus-Levitin (ML) theorem to establish a quantum speed limit. Specifically, we consider the weakest boundary of this limit, as elucidated by Niklas Hörnedal et al. \cite{hornedal2023margolus},
\begin{lemma}\label{lemma}
    Given the evolution time $\tau$ between two quantum states $v_1$ and $v_2$, the fidelity threshold $\delta_{\text{fid}}$, and the governing Hamiltonian $H_0$ that induces the evolution, the lower bound of $\tau$ is constrained by the following inequality $\tau \geq \tau_{\text{weakest}}(\delta_{\text{fid}})$, where $\tau_{\text{weakest}}(\delta_{\text{fid}})= \frac{4 \arccos^2{\sqrt{\delta_{\text{fid}}}}}{\beta\pi^2 \langle H_0 - \sigma_{\text{min}} \rangle}$.
\end{lemma}
Here $\sigma_{\text{min}}$ is the smallest eigenvalue of the Hamiltonian $H_0$ and $\beta\approx 0.724$.

According to Lemma \ref{lemma} and guided by our central clustering criterion, which ensures that the geodesic distance between the parameterized state $\widehat{\psi}_j(\theta)$ and the true state $\psi_j$ remains below a certain threshold $\delta$, we establish the following theorem \ref{the:tau_save}. 
\begin{theorem}\label{the:tau_save}
    (Total Convergence Time Savings)
    When employing the clustering criterion $\|\widehat{\psi}_j(\theta) - \psi_j\|_F < \delta$ to relax the requirement for quantum step convergence, the quantum evolution time saved $\tau_{\text{save}}$ adheres to $\tau_{\text{save}} \geq \frac{4 \delta^2}{\beta\pi^2 \langle H(s) - \sigma_{\text{min}} \rangle}$.
\end{theorem}
Note that the relationship between the geodesic distance $\delta$ and the fidelity $\delta_{\text{fid}}$ for quantum states is $\delta = \arccos{\sqrt{\delta_{\text{fid}}}}$.

\subsection*{NISQ and FTQC implementation}\label{subsec2.2}
In this section, we explore the deployment of the framework introduced in this paper with PQCs as parameter representation across two quantum computing paradigms: NISQ devices and FTQC systems. For NISQ devices, we present a viable strategy utilizing PQCs, which is designed to iteratively approximate the ground state of the shift Hamiltonian $H(s)$. Regarding FTQC, we evaluate the feasibility of employing techniques such as the HHL algorithm and QSVT in harmony with our algorithm. The conditions necessary for the effective application of parameter-based methods are comprehensively discussed in the NISQ implementation Section. The FTQC implementation Section explores possible methods for our proposed algorithm under the FTQC paradigm.

\subsubsection*{NISQ implementation}\label{subsec2.2.1}
In this section, we discuss the requirements for designing PQCs to produce parameters that satisfy clustering conditions. Our goal is to distinguish eigenvalues within a given interval of $s$. To achieve this, two requirements need to be met. The first point is that PQC needs to be able to approximate the ground state of a series of Hamiltonians $H(s)$, which represent the eigenvector subspace. In other words, these PQCs must satisfy certain conditions, namely equations \eqref{s:i=j} and \eqref{s:i!=j}, as defined in the Parameter representation Section. The second point is that as the system scales, the number of circuit parameters that PQC meets does not increase exponentially while still maintaining a certain level of expressiveness.

For the first point, according to \cite{Funcke2021dimensional} and \cite{haug2021capacity}, the structure of a PQC affects its ability to express quantum states and the expression ability of a given quantum circuit can be quantified by the dimensions of the state manifold that the circuit can achieve. The dimensions of the circuit manifold should match the number of parameters of the quantum circuit and the circuit should be able to span the eigenvector subspace. Therefore, when the dimensions of the eigenvector subspace increase (e.g., when considering larger qubit systems), the circuit ansatz should be extended to maintain a state manifold that is not smaller than the eigenvector subspace, so that there exist PQCs that satisfy conditions \ref{s:i=j} and \ref{s:i!=j} and can successfully distinguish the eigenvalues in the corresponding interval.

For the second point, PQCs have a fixed gate structure and the number of free parameters scales polynomially with the qubit count, while the dimensionality of the vector space grows exponentially with the qubit count \cite{du2020expressive,PQC7}. Therefore, we can reduce the complexity from exponential to polynomial by analyzing the circuit parameters on a classical computer instead of directly analyzing the vector space. Since we do not need to calculate the kth excited state sequentially, the circuit depth is shallower. The main source of complexity of our algorithm is the evolution of $H(s)$ on quantum computers. 

Today,  we can use $(H-sI)^2$ to implement $H(s)$. It can be realized through Imaginary-Time Evolution (ITE) \cite{mcardle2019variational}, which makes the function $f(s,t) = \bra{\psi_0} \exp\{-(H-sI)^2t\} \ket{\psi_0}$ converge to the eigenvalue of $H$ closest to $s$, and the state $\ket{\psi(s,t)}$ converge to the corresponding eigenstate of $H$, where $\ket{\psi_0}$ is an initial state. Under this evolution, the computational complexity will mainly depend on the system size $O(n)$ and the Trotterization method used of $(H-sI)^2$.

\subsubsection*{FTQC implementation}\label{subsec2.2.2}
In this section, we discuss the feasibility of employing techniques such as the HHL algorithm and QSVT to harmonize our algorithm. First, in the era of FTQC, the construction of $H(s)=(H-sI)^{2}$ can be replaced by $(H-sI)^{-1}$ using HHL or QSVT to get the better speed of convergence near the eigenstates. Then, the preparation of the eigenstates can be either done by a variational method \cite{chiew2023scalable} or the inverse power method \cite{kyriienko2020quantum}. 

For the variational method, Chiew et al. using the HHL method minimizing the objective function $f(s) = \bra{\widehat{\psi}_0(\theta)} (H-sI)^{-1} \ket{\widehat{\psi}_0(\theta)}$, where $\ket{\widehat{\psi}_0(\theta)}$ is a variational initial state to find the optimal $\theta$ that satisfies the constraints with the runtime analysis. Such parameter $\theta$ can be used to distinguish different eigenstates and estimate the eigenvalues corresponding to $s$ according to our Theorem \ref{the:fid_req_clu}. Although the HHL with variational ansatz seems to be a good match for our method, we point out that with the introduction of the drift term $s$, the matrix $H(s)$ may exhibit a significantly elevated condition number. In such scenarios, QSVT emerges as a superior alternative to the HHL algorithm. QSVT adeptly adjusts the polynomial approximation applied to $H(s)$, thereby offering enhanced error control and increased stability of the solution. This strategic choice leverages the inherent flexibility of QSVT to manage high-condition-number matrices, ensuring a robust and reliable computational process. Additionally, the inverse approximation hardness of QSVT near zero can be mitigated by combining it with our method.

For the inverse power method, after implementing the QSVT to $H(s)=(H-sI)^{-1}$ enables the generation of quantum states closely approximating the ground state of the Hamiltonian $H(s)$. Subsequently, shadow tomography facilitates the development of a measurement protocol comprising a sequence of observables $\{O_1, O_2,\cdots, O_{poly(n)}\}$ \cite{king2024triply,flammia2011direct}. These observables are crafted to distinguish expectation values for orthogonal quantum states, which are distinctly separable. Provided that the chosen set of observables for shadow tomography is sufficiently comprehensive—potentially forming an operator space basis—one can establish bounds on the state distance. Moreover, if the statistical distance (e.g., Hellinger distance) between the expectation value distributions produced by the observables correlates with the state distance, then, according to Theorem \ref{the:fid_req_clu}, these expectation values can be used as features for clustering.


\subsection*{Numerical experiments}\label{subsec2.4}
The main goal of this section is to validate our approach of clustering the quantum circuit parameters to distinguish the range of Hamiltonian eigenvalues, as proposed in the previous section. We use numerical simulations to demonstrate the effectiveness and scalability of our method on two different models: the 1-D Heisenberg model and the LiH molecular model. On the one hand, we show how our method can obtain the rough eigenspectrum of the Hamiltonian on both noiseless and noisy quantum devices by simulating the 1-D Heisenberg model. Secondly, we show the feasibility of our approach as the system scales up.
On the other hand, we show the numerical results of our method on the LiH molecular model, which reveals the application potential of our method in comparing the band gap differences under different internuclear separations.
\subsubsection*{One-dimensional Heisenberg model} \label{subsec2.5.1}

Firstly we discuss the noiseless case. The 1-D Heisenberg model is $$H = -\frac{1}{2}\sum_{j=1}^{N-1}(J_x\sigma_j^x\sigma_{j+1}^x + J_y\sigma_j^y\sigma_{j+1}^y + J_z\sigma_j^z\sigma_{j+1}^z + h\sigma_j^z).$$

Let $N=4$, $J_x=J_y=0.5$, $J_z=0.6$ and $h=1$, randomly generate the initial $\theta$ to run our algorithm. Then use the Hopkins statistic \cite{adolfsson2019cluster} on the obtained parameters data to validate clusters. Select the appropriate number of K-means clusters combined with the silhouette coefficient, delete outliers based on the Inter-Quartile Range, and give the clustering results. Since our algorithm is affected by the initial value, we generate the initial $\theta$ multiple times and compare the results of clustering of K-means 50 times with the same k-value. Select samples that can generate more classes and have larger silhouette coefficient values for analysis.

Before clustering, we first detect whether there is a cluster structure in the data distribution. If the data is essentially random, the results of clustering will be meaningless. To test this, we can observe whether the clustering error changes monotonically as the number of clusters increases. If the data is basically random, that is, there is no cluster structure, then the clustering error should change less significantly as the number of clusters increases. However, if the data has some underlying structure, the clustering error should decrease rapidly at first and then level off as the number of clusters reaches the optimal value. This article uses the Hopkins() function in R \cite{yilan2015clustertend} to implement the Hopkins statistic. The Hopkins statistic measures how clusterable a data set is by comparing its variance with the variance of a randomly generated data set of the same size and dimension. The Hopkins statistic ranges from 0 to 1, where values close to 1 indicate high clustering and values close to 0 indicate low clustering. We performed the Hopkins statistic 100 times on the parametric data and obtained average values of 0.998 and 0.981 respectively, which are very close to 1. We also calculated the p-value of the Hopkins statistic, which tests the null hypothesis that the data are not clusterable. A low p-value ($p<0.05$) rejects the null hypothesis and indicates that the data is more clusterable than random data, whereas a high p-value ($p\geq 0.05$) fails to reject the null hypothesis and indicates that the data is not significantly more clusterable than random data. In our case, the p-value for the Hopkins statistic is approximately zero, which means that we can confidently reject the null hypothesis and conclude that our data are very clusterable at the $5\%$ significance level. This also supports our theory that the parameter data in our algorithm has some underlying structure that can be revealed through clustering methods.

\begin{figure}[!htp]
\centering
\includegraphics[width=0.45\textwidth]{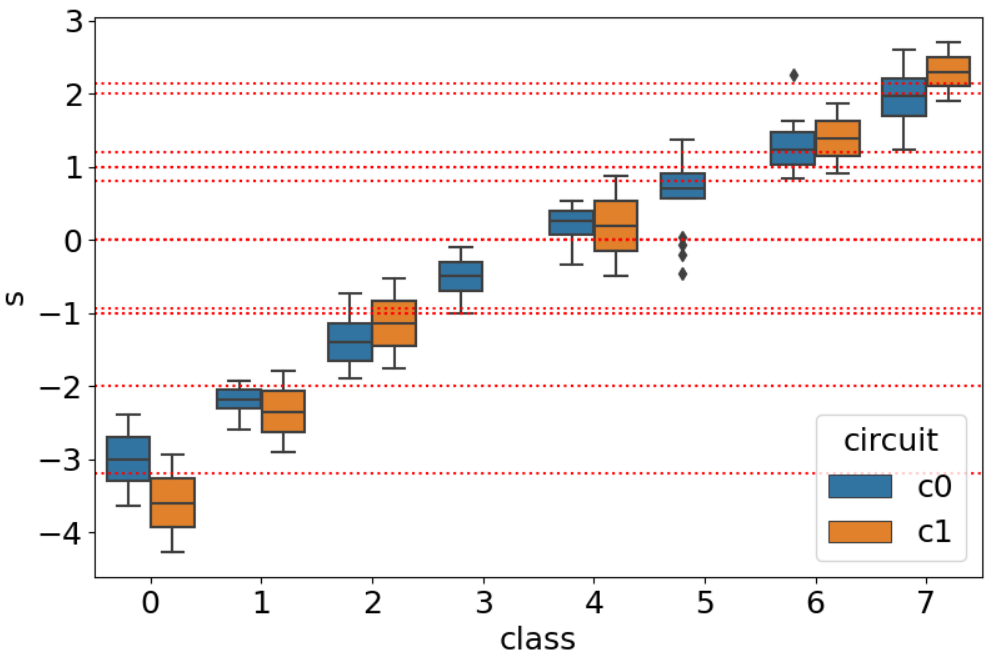}
\caption{Clustering result of a 4 qubits Heisenberg model. The x-axis represents the clustering category of the parameter, the y-axis represents the drift s, and the red dotted line represents the exact eigenspectrum.}
\label{fig:noiseless_h1}
\end{figure}

\begin{figure}[!htp]
\centering
\includegraphics[width=0.45\textwidth]{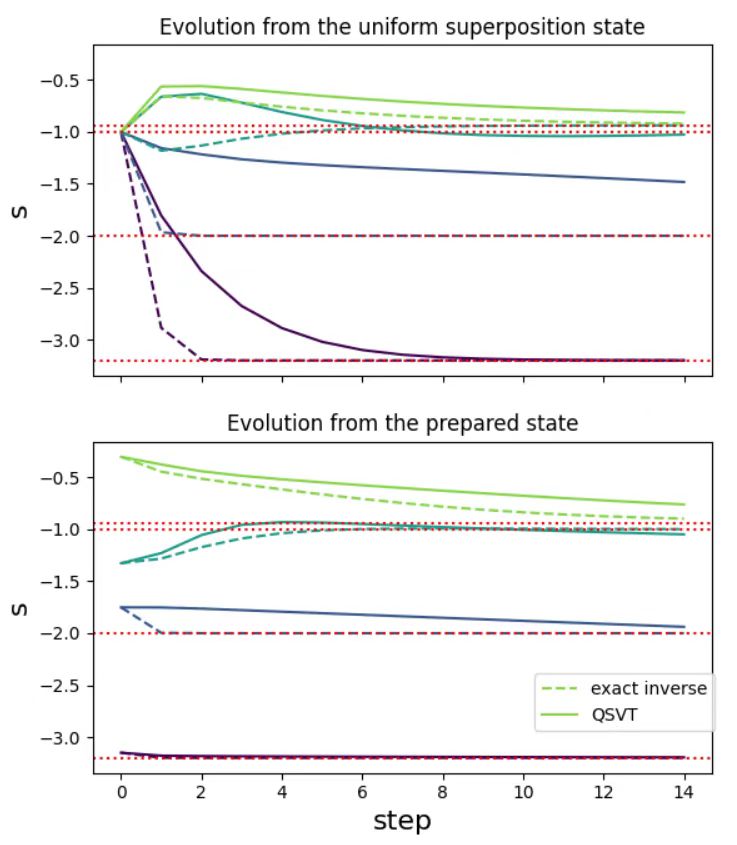}
\caption{The first and second subfigures illustrate the estimated energies obtained by applying the quantum inverse power method. The first subfigure uses the exact calculation of $H(s) = (H - sI)^{-1}$, while the second subfigure employs the polynomial approximation of $H(s)$ implemented via the QSVT. These estimations are shown for two initial states: a uniform superposition and the median state prepared as depicted in Fig. \ref{fig:noiseless_h1}. The x-axis represents the Iteration steps, the y-axis represents the drift s, and the red dotted line represents the exact eigenspectrum.}
\label{fig:noiseless_h1_final}
\end{figure}
Fig. \ref{fig:noiseless_h1} shows the clustering results after 25 steps of the imaginary time evolution of the Hamiltonian by given different circuit ansatz. The s corresponding to the median line in the box plot is our estimated eigenvalue. Circuit $c0$ is the circuit 11 that we selected based on Sukin Sim's results \cite{sim2019expressibility}, which is excellent in expressibility and entanglement capability with additional layers and has reasonable circuit costs. The circuit $c1$ is a circuit we designed that contains only rotation y gates and control rotation y gates with six parameters.

From Fig. \ref{fig:noiseless_h1}, we can see that the distribution of $s$ corresponding to parameters classified into the same category matches the distribution of the eigenspectrum. The ability of analysis depends on the relationship between the gap of eigenvalues and the step size of $s$. Our algorithm can accurately determine the interval containing the eigenvalues, but using the median to represent the interval may introduce errors due to uneven distribution of eigenvalues. When a more accurate energy spectrum is required, we can refine the initial rough characteristic spectrum by incorporating the median $s$ and the corresponding prepared state. This can be done using either the imaginary time evolution based on $H(s) = (H - sI)^2$ or the quantum inverse power method based on $H(s) = (H - sI)^{-1}$ to achieve a more precise estimate. Fig. \ref{fig:noiseless_h1_final} presents the evolution results after employing QSVT to implement the polynomial approximation of $H(s) = (H - sI)^{-1}$. It demonstrates that initiating the evolution with the quantum state corresponding to the median of the group leads to faster iterations compared to using a general state, such as a uniform superposition state. Additionally, Fig. \ref{fig:noiseless_h1_final} shows a scenario where the uniform superposition state is very close to the corresponding excited state. Furthermore, Fig. \ref{fig:noiseless_h1} indicates that different circuit designs affect the accuracy of the estimation.

Then we use the noise module in Qiskit Aer \cite{Qiskit} to construct a basic depolarizing error noise model. The process is the same as described above. In our example, we assume that, for all single-bit gates, the depolarization noise is 0.001. For the double quantum gate $cx$, the depolarization noise increases from 0.005, 0.010 to 0.030. To observe the classification effect when the number of evolution steps is different in the presence of noise, we evolved the algorithm for 25 steps and recorded the parameters of steps 5, 10 to 25 for clustering. We represent the error using the absolute value of the difference between the exact value of the eigenvalue and the value obtained by our method. We perform a Mann-Kendall Trend Test in R using the MannKendall() function from the Kendall library \cite{mktest}. The results are shown below.

\begin{figure}[!htp]
\centering
\includegraphics[width=0.5\textwidth]{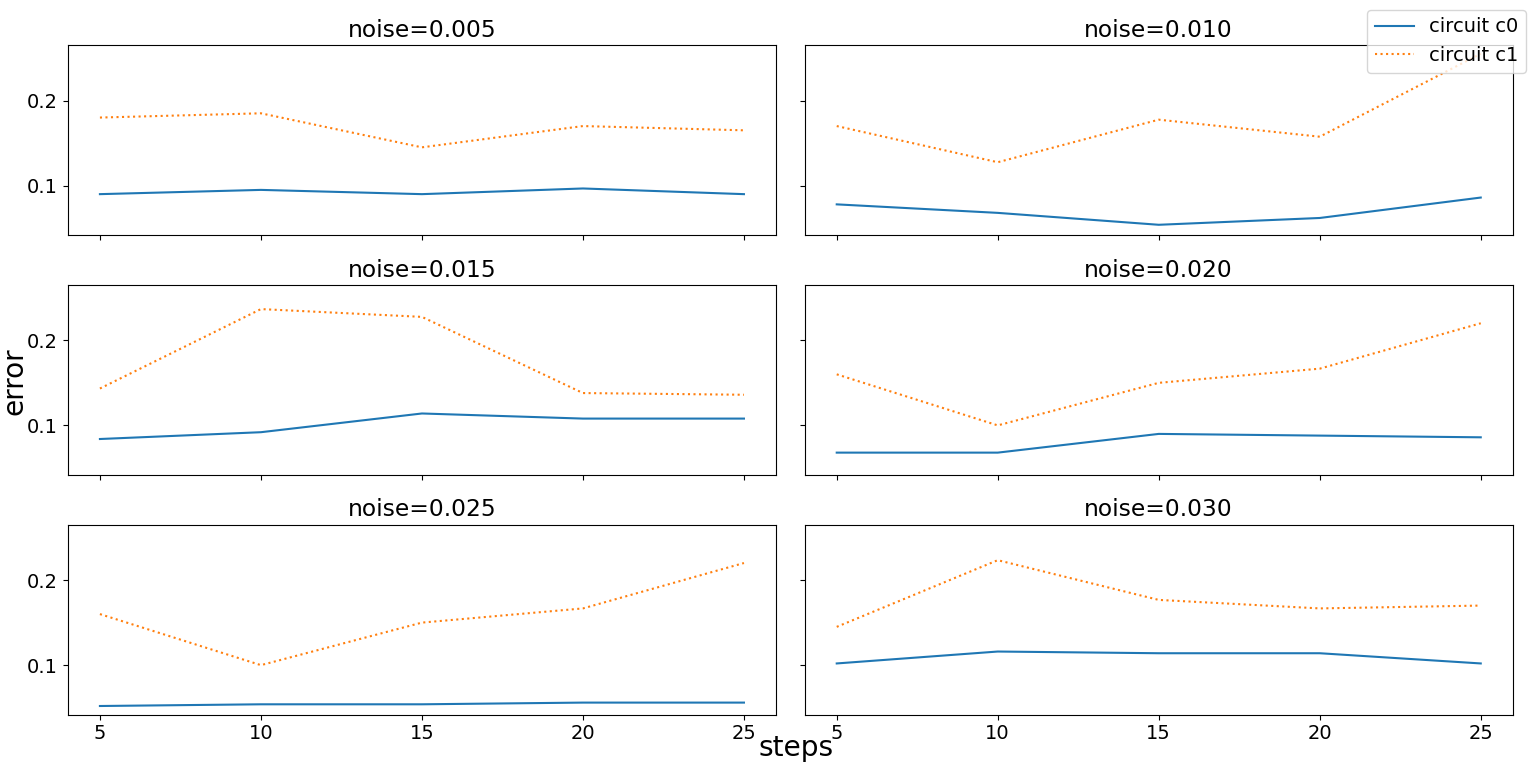}
\caption{Average error under different noise.}\label{fig:noise_h1}
\end{figure}

Fig. \ref{fig:noise_h1} shows the error results when the noise increases from 0.005 to 0.03 under different evolution steps and circuits. We can find that the errors are affected by ansatz, and using circuits with high expressivity and entanglement can reduce the errors. In addition, we also tested the situation when the error continued to increase to 0.05. We found that when the noise is greater than 0.03, running our algorithm using circuit $c1$ fails to cluster. According to the error when the noise increases from 0 to 0.05 or 0.03 at different evolution steps (the maximum noise we use is 0.05 for circuit $c0$, and 0.03 for $c1$), we obtain the corresponding Mann-Kendall trend test result. The Mann-Kendall trend test uses Kendall’s tau as the statistic, to see if there is a certain trend in the error as the noise increases. The Mann-Kendall trend test is a nonparametric test that does not assume any particular distribution of the data and relies only on the relative ranking of observations. Since all the p-values we get are much greater than 0.05, we can be sure that the null hypothesis cannot be rejected. As the noise increases, there is no significant trend in errors. In other words, our circuit has excellent anti-noise ability and is robust. 


Next, we show the scalability of our model as the system scales. We aim to numerically demonstrate the feasibility of our algorithmic theoretical framework at different circuit widths. We apply our algorithm to various quantum datasets with different numbers of qubits ($n=6$ to $12$, $J_x=J_y=0.5$, $J_z=0.6$, and $h=1$). For each Hamiltonian $H$ and its corresponding circuit width, the value range of s includes the range of the first four eigenvalues of $H$. We then calculate the clustering results of the circuit parameters, and the results are shown in Table \ref{tab:scaling_h1}. The experimental results indicate that under the same circuit ansatz, as the number of qubits increases, the error remains relatively stable when calculating the first four feature values. This implies that the observation results and insights obtained with smaller qubit numbers can be extrapolated to infer the performance of our algorithm under the same circuit ansatz with a larger number of qubits. We can imply that our algorithm is scalable and applicable to a model with a large number of qubits.

\begin{table}[!htp]
\begin{tabular}{@{}ccc@{}}
\toprule
number of qubits & average error & number of parameters \\ \midrule
6 & 0.061 & 10 \\
8 & 0.099 & 14 \\
10 & 0.078 & 18 \\
12 & 0.065 & 22 \\ \bottomrule
\end{tabular}
\caption {The average error of different circuits. We use the absolute error of theoretical value and estimation value to represent errors. The average error is the average error of the first four feature values. The theoretical value is the median of the exact value range.} \label{tab:scaling_h1} 
\end{table}
Note that we use the absolute error of theoretical value and estimation value to represent errors. The average error is the average error of estimating the first four eigenvalues. The theoretical value is the median of the corresponding range of exact eigenvalue. We use theoretical values rather than exact values for the following reason: $H$ in Fig. \ref{fig:noise_h1} is fixed, so the exact eigenspectrum can be used as a fixed reference to observe errors. However, $H$ changes with n, which means that their eigenspectra are different. Therefore, the difference between the corresponding median and the exact value is not fixed, and the exact values of the eigenvalues are not a fixed reference. To measure the trend of errors when the number of qubits increases, we must calculate the theoretical value of the algorithm and measure the error using the theoretical value and estimation value.

\subsubsection*{LiH molecule} \label{subsec2.5.2}
In this section, we simulate the Hamiltonian of diatomic LiH molecules with varying internuclear distances using a 12-qubit quantum circuit. The energy estimation procedure is similar to the one described in the previous section. However, we find that the 12-qubit ansatz we use has a slow convergence rate for this problem. To overcome this challenge, we used some prior knowledge about the Hamiltonian properties as a guide for choosing the initial values of the parameters. The prior knowledge is based on the fact that the chemical electronic Hamiltonian with a similar chemical structure will have a similar eigenstate wavefunction. Therefore, we first solve a Hamiltonian with a given internuclear separation and then use the converged parameters obtained by our method as the initial values for another Hamiltonian with a close internuclear separation. We experimentally demonstrate that by initializing subsequent Hamiltonians with parameter values corresponding to the cluster centers solved for the Hamiltonian associated with previous internuclear separations—starting from the Hamiltonian corresponding to the largest separations—we improve the convergence rate by an order of magnitude.

\begin{figure}[!htp]
\centering
\includegraphics[width=0.5\textwidth]{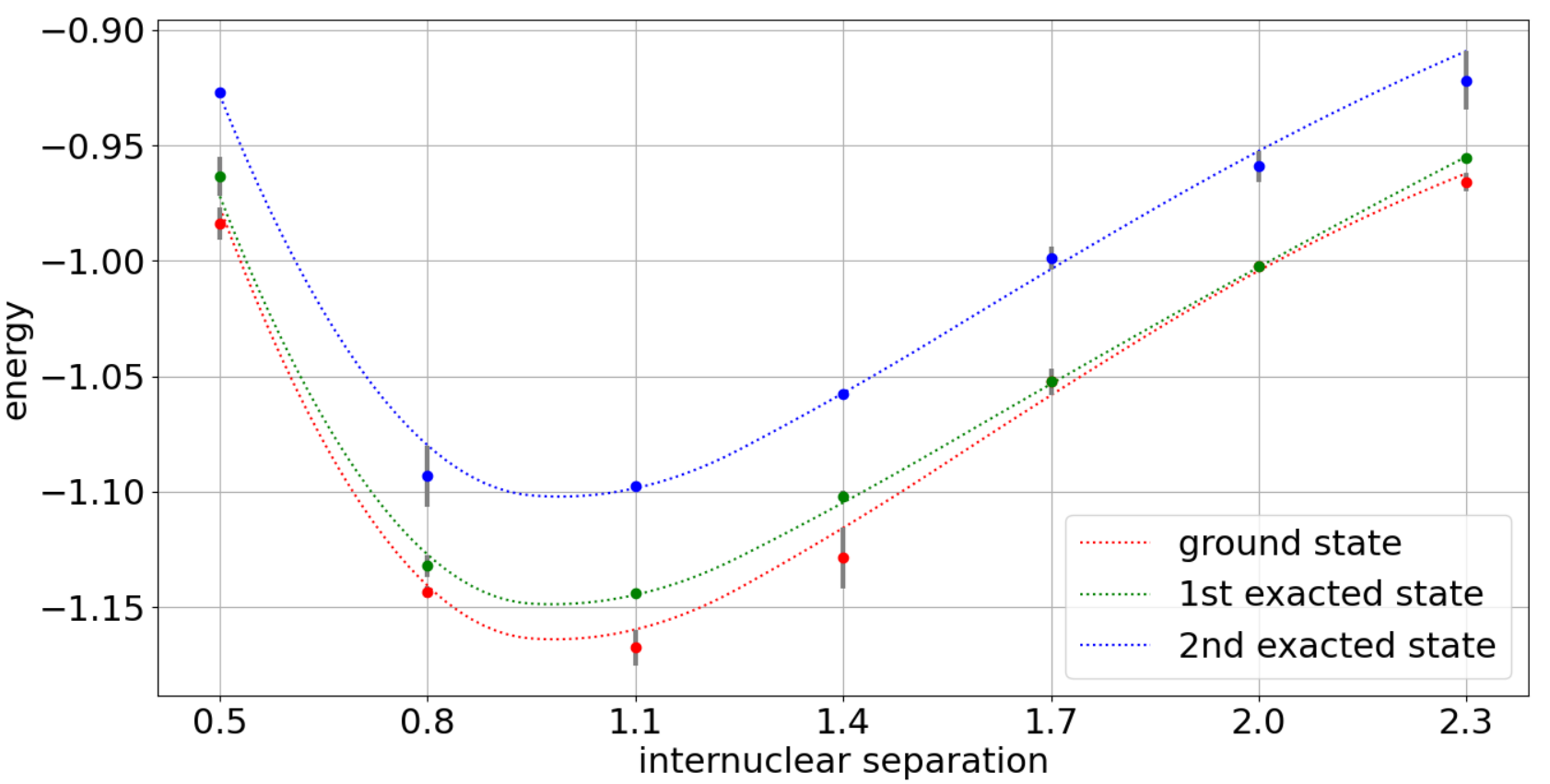}
\caption{The energy of the ground state and lower excited states versus internuclear separation for LiH molecules. In this figure, the exact energy values are shown by the dashed line, while our method estimates the energy using the same ansatz as the scatter points. This figure shows the absolute error bars of our estimates, as indicated by the gray lines.}
\label{fig:molecular}
\end{figure}

Fig. \ref{fig:molecular} shows the results of our method compared with the exact energy values. 
When the internal separation is 1.7 and 2.0, our algorithm cannot distinguish between the ground state and the first excited state. This is due to the step size of the drift $s$. Other sources of error include insufficient ansatz and a systematic error caused by using the median as an estimator.

\subsubsection*{Error analysis}
According to our algorithm and the experimental results, we found that whether the eigenspectrum can be found accurately depends on two factors: (1) whether the step size of the drift is appropriate; and (2) whether the setting of the ansatz is appropriate.

\begin{figure}[!htp]
\centering
\includegraphics[width=0.5\textwidth]{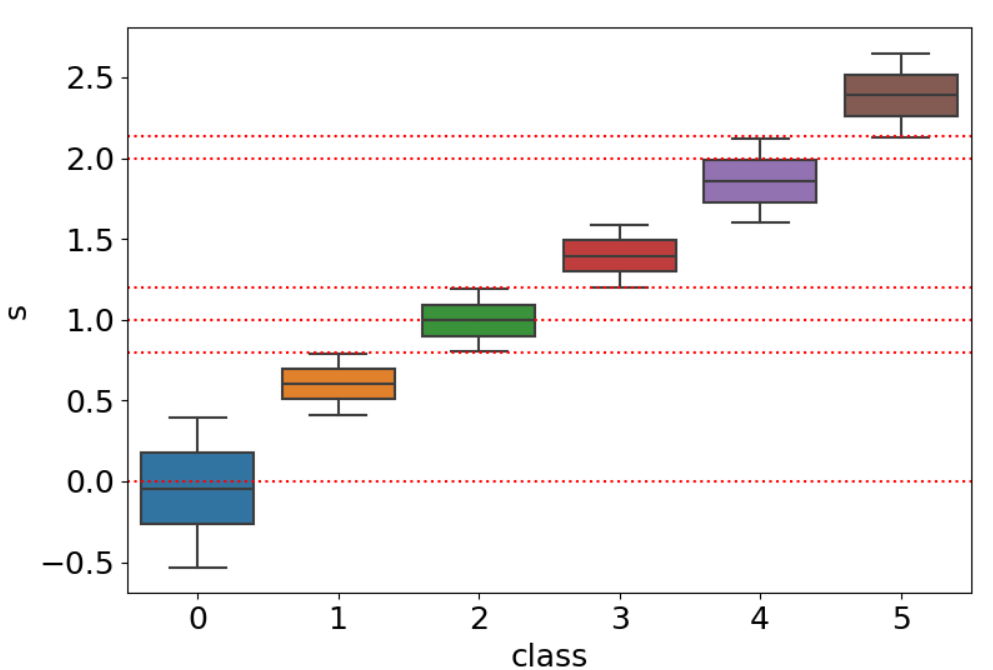}
\caption{Clustering results using a smaller step size of drift.}
\label{fig:de_noiseless_h1}
\end{figure}

For the first factor, we can more carefully filter the eigenvalues of the interval of interest by reducing the step size by the drift. The results are shown in Fig. \ref{fig:de_noiseless_h1}. Compared to the results of circuit $c0$ in Fig. \ref{fig:noiseless_h1}, we can see that by reducing the step size of drift $s$, we can distinguish closer eigenvalues. However, the same-sized interval has more values of $s$, and we need to do more evolutions for the operator $(H-sI)^2$. For the second factor, as we discussed in the previous section, different ansatz have different expressive capabilities for the same Hamiltonian. Using multiple ansatz to average their output can help reduce errors.

\section*{Discussion}\label{sec4}
In the present work, we have proposed a quantum-classical hybrid framework that leverages the complementary strengths of quantum state evolution and classical data processing. This framework is designed to overcome the computational challenges of both quantum and classical simulation methods based on two key theorems. First, Theorem \ref{the:fid_req_clu} demonstrates that quantum circuit parameters can effectively differentiate Hamiltonian eigenvectors under specific conditions. Second, Theorem \ref{the:tau_save} suggests that classically clustering quantum states relaxes convergence requirements. To fulfill the framework, we proposed different implementations in the era of NISQ and FTQC, we then implemented numerical simulations to validate our approach's effectiveness and scalability across different models in the NISQ setting. We first confirmed the high clustering of the data at the $5\%$ significance level based on Hopkins statistics, supporting the structured properties of the parameter data proposed in our Theorem \ref{the:fid_req_clu}. Then, Numerical experiments show that our approach can quickly estimate the eigenspectrum of the one-dimensional Heisenberg model regardless of the presence or absence of noise. That is, we can complete clustering using very relaxed convergence conditions and obtain a rough eigenspectrum. The Mann-Kendall trend test confirmed the algorithm's robustness, as it displayed no significant trend in errors with increasing noise levels. Finally, our algorithm’s performance remains relatively constant as the number of qubits increases, implying scalability. These results collectively underscore the hybrid algorithm's potential as a scalable and noise-resistant solution. In conclusion, in comparison to existing hybrid eigenspectrum solvers, which typically require either deeper quantum circuits, an extensive number of measurements, or a step-by-step approach to obtaining the eigenspectrum, our framework presents a significant advancement. As we look to the horizon of quantum computing research, several compelling questions warrant further investigation. First, will the adoption of quantum clustering over classical clustering yield quantum advantages?
Second, is there a more efficacious method for constructing the parameter representation of quantum states within the FTQC framework? Finally,  What impact will the relaxed fidelity requirements, as a result of clustering, have on gate complexity?
These questions not only highlight the potential for significant strides in the field but also open avenues for future scholarly inquiry.

\section*{Method}
\subsection*{A dynamic process that satisfies the conditions of the Result Section} \label{secA2.1}
We aim to use a parametric quantum circuit to approximate the ground states of a series of Hamiltonians $H(s)=(H-sI)^2$, which is the kth excited state of the Hamiltonian $H\in \mathcal{H}^m$ that satisfies $\lambda_k=\arg\min_{\lambda_i} |\lambda_i-s|$. Here, $\lambda_i$, $i=1,2\cdots,m$ and $m=2^n$, represent the eigenvalues of $H$. In this way, we can transfer the initial state to different excited states of the Hamiltonian $H$. Various variational algorithms can find the ground state of a Hamiltonian, such as the Variational Quantum Eigensolver (VQE) and the VITE. However, ITE has a unique advantage over other methods: it exponentially suppresses the excited states and enhances the ground state, as long as the initial state has some overlap with the ground state. Therefore, ITE does not get trapped at a local minimum but converges with the global minimum, which is ground-state energy. VITE is a hybrid quantum-classical algorithm that approximates ITE using a parametric quantum circuit. VITE can find accurate estimates of ground state energy and state with fewer circuit parameters than other gradient-based methods, such as VQE. Although VITE can also encounter local minima if the expressivity of the parametric quantum circuit is insufficient to capture the ground state, high accuracy can be achieved with proper choices of the variational ansatz and the initial state \cite{stokes2020quantum,alam2023solving,mcardle2019variational}. Considering the faster convergence rate of ITE and the tendency of gradient-based methods to fall into local minima, we believe that VITE requires fewer measurements than gradient descent to obtain an approximate ground state of $(H-sI)^2$ that satisfies the error bounds, despite needing more measurements to construct the McLachlan distance matrix. Consequently, in this work, we delineate the implementation of the Hamiltonian evolution $$ H(s) = (H - sI)^2 $$ employing VITE, thereby potentially enhancing the efficiency of ground state approximation.

For a given Hermitian matrix $H\in \mathcal{H}^m$ and an initial state $v_0$, set
\begin{equation}\label{A(t)1}
    A(s,t)=\exp\{-(H-sI)^2t\},
\end{equation}
\begin{equation}\label{B(t)v1}
    B(s,t)v_0=\frac{A(s,t)v_0}{||A(s,t)v_0||_2},
\end{equation}
\begin{equation}\label{f(t)1}
    f(s,t)=(B(s,t)v_0)^\dagger H (B(s,t)v_0).
\end{equation}

We can deduce the convergence of $f$ according to the following theorem.
\begin{theorem}\label{the:convergence}
    For any given scalar $s \neq \frac{\lambda_i+\lambda_{i+1}}{2}$, $i=1,...,m-1$, we have $$\lim_{t\to \infty} f(t)=\arg\min_{\lambda_i} |\lambda_i-s|.$$ 
\end{theorem}

\noindent\textit{Proof}: Suppose that 
$\lambda_1<\lambda_2<...<\lambda_m$. Since $H$ is Hermitian, there exists a unitary matrix $U$ and a diagonal matrix $\Lambda={\rm diag}(\lambda_1,\dots,\lambda_m)$ such that $H=U^\dagger \Lambda U$. It follows that
\begin{align}\label{A(t)1:U}
    A(s,t)=U^\dagger\exp\{-(\Lambda-sI)^2t\}U
\end{align}
and 
\begin{equation} B(s,t)v=\frac{U^\dagger\exp\{-(\Lambda-sI)^2t\}Uv_0}{\sqrt{\sum\limits_{i=1}^N e^{2d_it}|v_i|^2}}
\end{equation}
where $d_i=-(\lambda_i-s)^2$ and $v_i$ is the $i$-th component of the vector $Uv_0$.

If $x_j$, $j=1,\dots,m$, is the $j$-th component of $x:=UB(s,t)v_0$, then   
$$x_j=\left(\sum\limits_{i=1}^m e^{2(d_i-d_j)t}|v_i|^2\right)^{-1/2}v_j.$$
Note that as $t \to \infty $, 
$$x_j\rightarrow\begin{cases} 0 &{\rm if}\ d_j \neq \max_i \{d_i\};\\
\frac{v_j}{|v_j|}    &{\rm if}\ d_j = \max_i \{d_i\}.
\end{cases}$$

Now according to the hypothesis of $s$, there is a unique $1\leq k\leq m$ such that 
$$\lambda_k=\arg\max_{\lambda_i} \{d_i\}=\arg\min_{\lambda_i} |\lambda_i-s|.$$
It follows that 
$$\lim_{t \to \infty}x= (0,\cdots,\frac{v_k}{|v_k|},\cdots,0)^T,$$
and according to the defining equation \eqref{f(t)1}, 
\begin{align*}
    \lim_{t \to \infty} f(s,t)&=\lim_{t \to \infty}(UB(s,t)v_0)^\dagger \Lambda (UB(s,t)v_0) \\
    &= \lim_{t \to \infty} x^\dagger \Lambda x=\lambda_k.
\end{align*}

\subsection*{Clustering and circuit design strategy}

We first discuss clustering methods, data normalization, and the choice of the number of clusters. We use the kmeans() function from sklearn \cite{scikit-learn} to cluster the circuit parameters based on their angle values. To normalize the data, we convert the angle values to real and imaginary parts using $\exp{i\theta}$, resulting in two features for each parameter. To avoid local minima and speed up convergence, we use the k-means++ initialization scheme, which samples the initial cluster centroids based on the empirical probability distribution of the point's contribution to the overall inertia. We also run the k-means algorithm multiple times with different centroid initializations and select the best result. To determine the optimal number of clusters, we use the silhouette coefficient, which measures the separation and compactness of the clusters. This is because different clustering algorithms will output different results given a preset number of clusters, in an unsupervised manner, we can evaluate the clustering effect by examining the separation of clusters and the compactness of clusters.

Next, we describe some details of the circuit design when expanding the system. The experimental results show that the better-performing ansatz for the 1-D Heisenberg model with 4 qubits is $c0$. This ansatz has $4(n-1)L$ parameters, $(n-1)L$ two-qubit operations, and $6L$ circuit depth, where $n$ is the number of qubits and $L$ is the number of circuit layers. To observe the performance of the algorithm as $n$ increases, this ansatz is used. Furthermore, the Fisher score \cite{tang2014feature} is used to calculate the contribution of each circuit parameter based on the obtained eigenspectrum and parameter data. The Fisher score sorting and comparison reveal that the parameters corresponding to the rotation z gate contribute approximately 30\% to distinguishing the first 3 eigenvalues and are ranked lower. Therefore, we try to build a lower-cost circuit ansatz $\hat{c}0$ based on $c0$ by removing the parameters of the rotating z gate. The effects of the circuit on numerical results before and after parameter deletion are as follows.

\begin{table}[!htp]
\caption {Average error under different circuit ansatz.} \label{tab:ansatz_scaling} 
\begin{tabular}{@{}ccccc@{}}
\toprule
circuit & n & \begin{tabular}[c]{@{}c@{}}average\\ error\end{tabular} & \begin{tabular}[c]{@{}c@{}}number of\\  paramters\end{tabular} & \begin{tabular}[c]{@{}c@{}}circuit\\ depth\end{tabular} \\ \midrule
\multicolumn{1}{|c|}{\multirow{4}{*}{c0}} & 6 & 0.109 & 20 & 6 \\
\multicolumn{1}{|c|}{} & 8 & 0.158 & 28 & 6 \\
\multicolumn{1}{|c|}{} & 10 & 0.159 & 36 & 6 \\
\multicolumn{1}{|c|}{} & 12 & 0.062 & 44 & 6 \\ \cmidrule(r){1-1}
\multicolumn{1}{|c|}{\multirow{4}{*}{$\hat{c}0$}} & 6 & 0.069 & 10 & 4 \\
\multicolumn{1}{|c|}{} & 8 & 0.103 & 14 & 4 \\
\multicolumn{1}{|c|}{} & 10 & 0.062 & 18 & 4 \\
\multicolumn{1}{|c|}{} & 12 & 0.074 & 22 & 4 \\ \bottomrule
\end{tabular}
\end{table}

Based on the experimental results in Table \ref{tab:ansatz_scaling}, circuit $\hat{c}0$ was chosen for the scaling experiment of H in numerical experiments. This is because circuit $\hat{c}0$ may have better performance and lower cost for the current $H$, as shown by the data.

\bmhead{Acknowledgments}

AH gratefully acknowledges the sponsorship from the Research Grants Council of the Hong Kong Special Administrative Region, China (Project No. CityU 11200120), City University of Hong Kong (Project No. 7005615, 7006103), CityU Seed Fund in Microelectronics (Project No. 9229135).

\bibliography{sn-bibliography}



\end{document}